\documentclass[journal]{IEEEtran}
\usepackage{amsmath}
\usepackage{amssymb}
\usepackage{setspace}
\usepackage{epsfig}
\usepackage{epstopdf}
\usepackage{rotating}
\usepackage{color}
\usepackage{colortbl}
\usepackage{subfigure}

\newcommand{\vt}[1]{\ensuremath{\mathbf{#1}}} 
\newcommand{\lt}[1]{\ensuremath{\text{#1}}} 
\newcommand{\unitx}{\ensuremath{\vt{u}_x}}
\newcommand{\unity}{\ensuremath{\vt{u}_y}}
\newcommand{\unitz}{\ensuremath{\vt{u}_z}}

\newcommand{\unitr}{\ensuremath{\vt{u}_r}}

\newcommand{\unit}{\ensuremath{\vt{u}}}
\newcommand{\er}{\ensuremath{\epsilon_\lt{r}}}
\newcommand{\ur}{\ensuremath{\mu_\lt{r}}}
\newcommand{\aee}{\ensuremath{\overline{\overline{\alpha}}_\lt{ee}}}
\newcommand{\aem}{\ensuremath{\overline{\overline{\alpha}}_\lt{em}}}
\newcommand{\ame}{\ensuremath{\overline{\overline{\alpha}}_\lt{me}}}
\newcommand{\amm}{\ensuremath{\overline{\overline{\alpha}}_\lt{mm}}}

\newcommand{\adyad}{\ensuremath{\overline{\overline{\alpha}}}}

\newcommand{\unitdyad}{\ensuremath{\overline{\overline{I}}}}
\newcommand{\Jdyad}{\ensuremath{\overline{\overline{J}}}}

\def\Re{{\rm Re\mit}}
\def\Im{{\rm Im\mit}}
\def\l#1{\label{eq:#1}}
\def\r#1{(\ref{eq:#1})}

\begin{document}

\title{Eliminating Electromagnetic Scattering from Small Particles}

\author{Joni  Vehmas, Younes Ra'di,~\IEEEmembership{Student Member,~IEEE}, Antti O. Karilainen,~\IEEEmembership{Member,~IEEE} and Sergei Tretyakov,~\IEEEmembership{Fellow,~IEEE}
\thanks{J.~Vehmas, Y. ~Ra'di and S.~A.~Tretyakov are with the Department of Radio Science and Engineering/SMARAD Center of Excellence, Aalto University, P.~O.~Box~13000, FI-00076 AALTO, Finland. Email: joni.vehmas@aalto.fi.}%
\thanks{A. O.~Karilainen is with Nokia Corporation, P. O. Box 226, FI-00045 Nokia Group, Finland.}}

\markboth{Vehmas \MakeLowercase{\textit{et al.}}: Eliminating Electromagnetic Scattering from Small Particles}%
{Vehmas \MakeLowercase{\textit{et al.}}: Eliminating Electromagnetic Scattering from Small Particles}
\maketitle

\begin{abstract}
\boldmath
This paper presents and discusses the conditions for zero
electromagnetic scattering by electrically small particles. We
consider the most general bi-anisotropic particles, characterized by
four dyadic polarizabilities and study the case of uniaxially
symmetric objects. Conditions for zero backward and forward
scattering are found for a general uniaxial bi-anisotropic particle
and specialized for all fundamental classes of bi-anisotropic
particles: omega, ``moving'', chiral, and Tellegen particles.
Possibility for zero total scattering is also discussed for
aforementioned cases. The scattering pattern and polarization of the
scattered wave are also determined for each particle class. In
particular, we analyze the interplay between different scattering
mechanisms and show that in some cases it is possible to compensate
scattering from a polarizable particle by appropriate
magneto-electric coupling. Examples of particles providing zero backscattering and zero forward scattering are presented and studied numerically.
\end{abstract}



\section{Introduction}
\label{sec:introduction}

\IEEEPARstart{E}{xamples} of finite-sized objects with zero backscattering are known
from the literature, a classical one being an object from an
isotropic material with equal relative constitutive parameters, $\ur
= \er$, and $\pi/2$ rotational symmetry when observed from the
incidence direction. \cite{Wagner1963, Weston1963}  The formal
condition for zero forward scattering in the case of a small
isotropic sphere was found in \cite{Kerker} and further discussed in \cite{Garcia-Camara, AluEng}. The last few years saw renewed
interest in ``invisible'' objects in the context of the concept of
cloaking (e.g., \cite{cloaking_alu,cloaking_pendry,cloaking_leonhardt}), where the
objects produce, in the ideal case, zero total scattered power. In
papers on cloaking, the zero-scattering property is achieved either
by using a material with such electromagnetic properties that the
generated scattered wave interferes destructively with the wave
scattered from the cloaked object (scattering-cancelation technique
\cite{cloaking_alu}) or by using inhomogeneous distributions of
materials with exotic electromagnetic properties to create a volume
inside which the incident wave cannot penetrate
(transformation-optics technique \cite{cloaking_pendry,
cloaking_leonhardt}). Alternatively, mesh-like objects can also be
cloaked by guiding the wave through the object via a
transmission-line network (transmission-line cloaking
\cite{Alitalo_inv}).

Electromagnetic scattering from electrically small particles is
determined by the lowest-order moments of induced current or
polarization, the electric and magnetic dipole moments. If the
particle material is an isotropic magnetodielectric, then the only
possibility to achieve zero backscattering is basically the trivial
case of equal relative parameters and proper symmetry of the
particle shape \cite{Wagner1963, Weston1963}. Similarly,
the only possible solution for zero forward scattering by a small
magnetodielectric sphere is the case found by Kerker \cite{Kerker}:
$\er=(4-\ur)/(2\ur+1)$, which in actuality corresponds only to
greatly diminished, but not identically zero, forward scattering \cite{AluEng,NV_JOSA}. 
Recently, it was shown theoretically \cite{NV_JOSA,NV_nano,Belov_JETP} and confirmed by microwave experiments \cite{Belov_APL,NV_nature} that 
electric and magnetic polarizabilities required for forming a zero-scattering Huygens 
pair of dipoles satisfying Kerker's conditions can be realized in a single dielectric sphere. In antenna engineering, several realizations of small antennas satisfying the zero-backscattering condition, analogous to the Kerker condition for an isotropic sphere, are known \cite{Underhill2000,Best_IWAT2010,Ziolkowski,Teemu}.

However, electrically small particles can exhibit bi-anisotropic magneto-electric coupling (e.g.,
~\cite{Serdyukov2001}), so that the electric moment of the particle
is generated by both electric and magnetic incident fields
(likewise, the magnetic moment is induced by both fields). We expect
that bi-anisotropy of particles can offer more possibilities for
controlling scattering, in particular, for realizing non-scattering
objects.

In this paper, we study zero-scattering conditions (in back and
forward directions) for general bi-anisotropic particles, including
both reciprocal and nonreciprocal polarizabilities and
field-coupling mechanisms. We assume that the particles are
electrically small, so that the dipole approximation is an
appropriate model of the particle response and write the relations
between the induced electric and magnetic dipole moments $\vt{p}$
and $\vt{m}$ in terms of four dyadic polarizabilities $\adyad$:
\setlength\arraycolsep{2pt}
\begin{eqnarray}
    \vt{p} & = & \adyad_\lt{ee} \cdot \vt{E}_\lt{inc} + \adyad_\lt{em} \cdot \vt{H}_\lt{inc},
    \label{eq:psimple} \\
    \vt{m} & = & \adyad_\lt{me} \cdot \vt{E}_\lt{inc} + \adyad_\lt{mm} \cdot \vt{H}_\lt{inc}.
    \label{eq:msimple}
\end{eqnarray}
Here,  $\vt{E}_\lt{inc}$ and $\vt{H}_\lt{inc}$ are the incident
fields. In earlier studies, conditions for zero backscattering from
isotropic chiral objects were found \cite{Uslenghi,Uslenghi2,Karilainen_chiral}, and a uniaxial chiral object with low backscattering was
investigated \cite{Karilainen_antenna}. In a recent paper
\cite{Lindell2009} zero-backscattering from bi-anisotropic objects
was mentioned in the context of scattering from objects made from
self-dual materials. It has been shown that in addition to the
$\pi/2$ rotational symmetry, self-dual nature of the bi-anisotropic
filling material is sufficient for zero backscattering. In terms of
the particle polarizabilities, self-duality conditions lead to the
requirements  $\amm  =  \eta_0^2 \aee$ and $\aem  =  -\ame$, where
$\eta_0$ is the free-space wave impedance \cite{Karilainen_chiral}.
As examples of self-dual objects, a so called DB sphere was studied
in \cite{Sihvola2009}, and a D'B' sphere and cube in \cite{Lindell2009}. Furthermore, it was shown in \cite{Lindell2009}
that the physical geometry of the object does not need to have the
$\pi/2$ rotational symmetry, as long as the geometrical asymmetry is
balanced by anisotropy of material response.

In the following, the scattering properties of a general uniaxial
bi-anisotropic particle are analyzed and the required conditions for
achieving zero backscattering, zero forward scattering as well as
zero total scattering (i.e., invisibility) are derived. The
fundamental classes of uniaxial bi-anisotropic particles (omega,
``moving'', chiral, and Tellegen particles) are studied
systematically, and the aforementioned zero scattering conditions for
each particle class are derived. Furthermore, two concrete particle designs, one corresponding to zero backscattering and the other one to zero forward scattering, are presented and studied numerically. In our analysis, the incident wave
is assumed to be linearly polarized. This study is relevant to the
design of absorbers, where the goal is to minimize reflection and/or
transmission and to the design of cloaks and low-scattering small
sensors, where the goal is to minimize the total scattering.
\section{Scattering from Electrically Small Bi-Anisotropic Particles}
\subsection{Polarizabilities and the Scattering Amplitude}
Scattered fields from a single electrically small bi-anisotropic
scatterer illuminated by an incident plane wave can be analyzed
based on the induced electric and magnetic dipole moments whose
relations to the incident electromagnetic fields are defined by
(\ref{eq:psimple}) and (\ref{eq:msimple}). For a reciprocal
particle, we have \cite{Serdyukov2001}:
\begin{equation}
    \aee = \aee^T, \quad \amm = \amm^T, \quad \lt{and} \quad \aem = -\ame^T
    \label{eq:reciprocal_adyads}
\end{equation}
and for a lossless particle
\begin{equation}
    \aee = \aee^\dag, \quad \amm = \amm^\dag, \quad \lt{and} \quad \aem = \ame^\dag,
    \label{eq:lossless_adyads}
\end{equation}
where $T$ denotes the transpose operation and $\dag$ is the Hermitian operator (transpose of the complex conjugate).
However, as we want to study the scattering from a particle, we should deal with the dynamic polarizabilities (the quasi-static model will be not adequate) and include scattering loss in the consideration, even for a ``lossless'' particle (particle with no absorption losses). In the last case, the polarizability dyadics are governed by the following set of equations following from the law of energy conservation \cite{Yatsenko}:
\begin{equation}
    \textnormal{Im}\{(\aee - \aem \cdot  \amm^{-1} \cdot \ame)^{-1}\} = \frac{k}{6 \pi \epsilon_0} \unitdyad,\l{5}
\end{equation}
\begin{equation}
    \textnormal{Im}\{(\amm - \ame \cdot \aee^{-1} \cdot \aem)^{-1}\} = \frac{k}{6 \pi \mu_0} \unitdyad,\l{6}
\end{equation}
\begin{equation}
    \textnormal{Re}\{(\aee - \aem \cdot \amm^{-1} \cdot \ame)^{-1} \cdot \aem \cdot \amm^{-1} \} = 0,\l{7}
\end{equation}
\begin{equation}
    \textnormal{Re}\{(\amm - \ame \cdot \aee^{-1} \cdot \aem)^{-1} \cdot \ame \cdot \aee^{-1}\} = 0.\l{8}
\end{equation}
Here, $k$ is the free-space wavenumber and $\unitdyad$ is  the
three-dimensional unit dyadic. In the limiting case of ideal planar
particles when the fields along the axial direction do not interact
with the particle, $2 \times 2$ dyadics can and should be used in
(\ref{eq:5})--(\ref{eq:8}) in order to avoid singular dyadics. For a
reciprocal particle, the last two equations are equivalent, and this
is true also for all non-absorbing nonreciprocal uniaxial particles
analyzed in this paper. It should be noted that due to the inclusion
of scattering losses all the polarizabilities can be complex
quantities whereas if the scattering losses are excluded for
lossless particles, $\aee$ and $\amm$ are always purely real while
$\aem$ and $\ame$ are purely imaginary for a reciprocal particle and
purely real for a nonreciprocal particle with $\aem = -\ame^\dag$.

The scattered electric far field from electric and magnetic dipole moments ($\vt{p}$ and $\vt{m}$) at the origin can be written as \cite{Lindell2009,Tai1971}
\begin{equation}
    \vt{E}_\lt{sca} = -\omega^2 \mu_0 \frac{e^{-jkr}}{4 \pi r} \vt{F}_\lt{sca},
    \label{eq:Esca}
\end{equation}
where the radiation vector $\vt{F}_\lt{sca}$ reads
\begin{equation}
    \vt{F}_\lt{sca} = \unitr \times (\unitr \times \vt{p}) + \unitr \times \frac{\vt{m}}{\eta_0},
    \label{eq:Fsca}
\end{equation}
$\unitr$ is unit vector in the scattering direction, $\eta_0$ is the wave impedance of free space, and $r$ is the radial distance from the origin. The normalized scattering pattern  $\vt{F}_\lt{sca, norm}$ for a given particle is given simply by $\vt{F}_\lt{sca}/\max\{|\vt{F}_\lt{sca}|\}$ as a function of the observation angles. Scattering directivity pattern, on the other hand, is defined as ratio of the power density the scatterer radiates in a given direction and the power density radiated by an ideal isotropic radiator radiating the same total power and is given by
\begin{equation}
    D = \frac{|\vt{F}_\lt{sca,norm}|^2}{\frac{1}{4 \pi}\int^{2 \pi}_0 \int^\pi_0 |\vt{F}_\lt{sca, norm}|^2 d\theta d\phi},
    \label{eq:Directivity}
\end{equation}
where $\theta$ and $\phi$ are the inclination and azimuthal angles of a spherical coordinate system, respectively.
%
%
\subsection{Zero Scattering under Plane-Wave Illumination: General Conditions}

Zero scattering in a certain direction for a given configuration of
exciting fields can be studied by demanding $\vt{F}_\lt{sca} = 0$ in
(\ref{eq:Fsca}) and using
the relation between the fields of the incident plane wave%
\begin{equation}
    \vt{H}_\lt{inc} = \frac{\vt{k}}{\omega\mu_0} \times \vt{E}_\lt{inc} = \frac{1}{\eta_0} \unit \times \vt{E}_\lt{inc},
    \label{eq:H_from_E}
\end{equation}
where $\unit$ is the incidence direction. This leads to the
condition
%
\begin{eqnarray} \nonumber
     \lefteqn{\unitr \times \aee \cdot \vt{E}_\lt{inc} + \unitr \times \frac{1}{\eta_0} \aem \cdot ( \unit \times  \vt{E}_\lt{inc} )} \\
         & & \ \ \ \ + \frac{1}{\eta_0} \ame \cdot \vt{E}_\lt{inc} + \frac{1}{\eta_0^2} \amm \cdot ( \unit \times  \vt{E}_\lt{inc}) = 0. \hspace{0.5cm}
     \label{eq:zero_sca_Fsca}
\end{eqnarray}
We can rewrite the term $\adyad \cdot (\unit \times \vt{E}_\lt{inc})$  as $(\adyad \times \unit) \cdot \vt{E}_\lt{inc}$,
which allows us to write the condition in a convenient form:
\begin{eqnarray} \nonumber
     (\unitr \times \aee  + \lefteqn{\unitr \times \frac{1}{\eta_0} \left( \aem \times \unit \right) + \frac{1}{\eta_0} \ame} \\
     & & \ \ \ \ + \frac{1}{\eta_0^2} \left( \amm \times \unit \right)) \cdot \vt{E}_\lt{inc} = 0. \hspace{0.5cm}
     \label{eq:zero_sca_Fsca3}
\end{eqnarray}
For the condition of zero backscattering we demand $\unitr = -\unit$ and get
\begin{equation}
    -\unit \times \aee  - \unit \times \frac{1}{\eta_0} \aem \times \unit + \frac{1}{\eta_0} \ame  + \frac{1}{\eta_0^2} \amm \times \unit = 0.
    \label{eq:zero_BS}
\end{equation}
To find the condition for zero forward scattering we demand in turn $\unitr = \unit$ and get
\begin{equation}
    \unit \times \aee  + \unit \times \frac{1}{\eta_0} \aem \times \unit + \frac{1}{\eta_0} \ame  + \frac{1}{\eta_0^2} \amm \times \unit = 0.
    \label{eq:zero_FS}
\end{equation}

\section{Particles under Study}
In this paper we study the zero-scattering conditions
(\ref{eq:zero_BS}) and (\ref{eq:zero_FS}) when the scatterer is a
uniaxial bi-anisotropic particle. The case of uniaxial (rotational)
symmetry is chosen in view of having zero scattering for any
polarization direction of the incident field. In view of potential
applications in periodical arrays at normal incidence we concentrate
on the effects of polarizabilities in the transverse plane and do
not include the axial polarizabilities in the analysis. In the most
general bi-anisotropic case, the transverse polarizabilities in
(\ref{eq:psimple}) and (\ref{eq:msimple}) take the forms
\setlength\arraycolsep{2pt}
\begin{eqnarray}
    \aee &=& \alpha_\lt{ee,S} \unitdyad_\lt{t} + \alpha_\lt{ee,A} \overline{\overline{J}}_\lt{t},
    \label{eq:aee_gen} \\
    \amm &=& \alpha_\lt{mm,S} \unitdyad_\lt{t} + \alpha_\lt{mm,A} \overline{\overline{J}}_\lt{t}.
    \label{eq:amm_gen} \\
    \aem &=& \alpha_\lt{em,S} \unitdyad_\lt{t} + \alpha_\lt{em,A} \overline{\overline{J}}_\lt{t},
    \label{eq:aem_gen} \\
    \ame &=& \alpha_\lt{me,S} \unitdyad_\lt{t} + \alpha_\lt{me,A} \overline{\overline{J}}_\lt{t},
    \label{eq:ame_gen}
\end{eqnarray}
where S and A refer, respectively, to the symmetric and
antisymmetric parts of the  corresponding dyadics,
$\unitdyad_\lt{t}$ is the transverse unit dyadic, that is, a dyadic
of the form ${\unitdyad_\lt{t}=\overline{\overline{I}} - \unitz
\unitz}$, ${\overline{\overline{J}}_\lt{t}=\unitz \times
\overline{\overline{I}}_\lt{t}}$ is the vector-product operator, and
$\unitz$ is the unit vector along the particle axis. In addition to
the general case, we will consider in detail the four canonical
bi-anisotropic particles as classified in \cite{Serdyukov2001, interactions}: omega, ``moving'', chiral, and
Tellegen particles. Omega and chiral particles are, based on
(\ref{eq:reciprocal_adyads}), reciprocal particles, whereas
``moving'' and Tellegen particles are nonreciprocal. These four
types are distinguished from each other solely by the values of the
symmetric and antisymmetric cross-coupling polarizability components as shown in Table~\ref{T:1}.
\begin{table}[h!]
\begin{center}
\caption{Different types of uniaxial bi-anisotropic particles
($\alpha$ can be any complex number).} \label{T:1}
\begin{tabular}{|l|c|c|c|c|}
\hline
\rowcolor[gray]{.9}
& $\alpha_\lt{em,S}$   & $\alpha_\lt{em,A}$    & $\alpha_\lt{me,S}$    & $\alpha_\lt{me,A}$ \\
\hline
\cellcolor[gray]{.9}
Omega particle & $0$ & $\alpha$ & $0$ & $\alpha$ \\
\hline
\cellcolor[gray]{.9}
``Moving'' particle & $0$ & $\alpha$ & $0$ & $-\alpha$ \\
\hline
\cellcolor[gray]{.9}
Chiral particle & $\alpha$ & $0$ & $-\alpha$ & $0$ \\
\hline
\cellcolor[gray]{.9}
Tellegen particle & $\alpha$ & $0$ & $\alpha$ & $0$ \\
\hline
\end{tabular}
\end{center}
\end{table}
%
\subsection{General Uniaxial Particle}

First, let us consider the most general uniaxial bi-anisotropic
particle by allowing all the polarizabilities, $\aee$, $\amm$,
$\aem$ and $\ame$, to have both symmetric and antisymmetric parts as
defined in (\ref{eq:aee_gen})--(\ref{eq:ame_gen}).
\subsubsection{Zero Backscattering}

In this general case, the
zero backscattering condition (\ref{eq:zero_BS}) reduces to
\begin{eqnarray}
\lefteqn{-\alpha_\lt{ee,S}(\unit \times \unitdyad_\lt{t})-\alpha_\lt{ee,A}(\unit \times \Jdyad_\lt{t})} \nonumber \\
& & \ \ \ -\frac{\alpha_\lt{em,S}}{\eta_0}(\unit \times \unitdyad_\lt{t} \times \unit) - \frac{\alpha_\lt{em,A}}{\eta_0}(\unit \times \Jdyad_\lt{t} \times \unit) \nonumber\\
& & \ \ \ +\frac{\alpha_\lt{me,S}}{\eta_0} \unitdyad_\lt{t} + \frac{\alpha_\lt{me,A}}{\eta_0} \Jdyad_\lt{t} \nonumber \\
& & \ \ \  +\frac{\alpha_\lt{mm,S}}{\eta_0^2}(\unitdyad_\lt{t} \times \unit) + \frac{\alpha_\lt{mm,A}}{\eta_0^2}(\Jdyad_\lt{t} \times \unit) = 0.
\label{eq:zerobackscat_gen}
\end{eqnarray}
By plugging a general directional  vector $\unit
=\sin{\theta}\cos{\phi}\unitx+\sin{\theta}\sin{\phi}\unity+\cos{\theta}\unitz=A\unitx+B\unity+C\unitz$
into (\ref{eq:zerobackscat_gen}), we can write a separate scalar
equation for each of the nine dyadic components. It turns out that
(\ref{eq:zerobackscat_gen}) for non-axial incidence can be satisfied
only for certain special cases, namely when we have
\begin{equation}
\alpha_\lt{em,S} = \alpha_\lt{me,S} = \alpha_\lt{ee,A} = \alpha_\lt{mm,A} = 0,
\label{eq:oblique_inc1}
\end{equation}
\begin{equation}
\alpha_\lt{ee,S} = \frac{\alpha_\lt{em,A}}{\eta} C = \frac{\alpha_\lt{me,A}}{\eta} \frac{1}{C} = -\frac{\alpha_\lt{mm,S}}{\eta^2}.
\label{eq:oblique_inc2}
\end{equation}
Notably even in this case, zero backscattering can be achieved  for
only one angle of incidence with given polarizabilities, and for the
incident direction orthogonal to the axis of the particle this is
not possible at all. Also, it was assumed that particle is ideally uniaxial meaning that the polarizability dyadics have no component along the particle axis ($\mathbf{u}_\lt{z} \mathbf{u}_\lt{z}$-component). If this component would be included, no solution could be found for oblique incidence.

For the axial incidence ($A= B = 0$, $C = \pm1$) using the identities
$\unit_\lt{z} \times \unitdyad_\lt{t} = \unitdyad_\lt{t} \times
\unit_\lt{z} = \Jdyad_\lt{t}$, $\unit_\lt{z} \times \unitdyad_\lt{t}
\times \unit_\lt{z} = - \unitdyad_\lt{t}$, and $\unit_\lt{z} \times
\Jdyad_\lt{t} \times \unit_\lt{z} = -\Jdyad_\lt{t}$, the dyadic
equation (\ref{eq:zerobackscat_gen}) reduces to two scalar
equations:
\begin{eqnarray}
\mp\alpha_\lt{ee,S}\pm\frac{\alpha_\lt{mm,S}}{\eta_0^2} + \frac{\alpha_\lt{em,A}}{\eta_0} + \frac{\alpha_\lt{me,A}}{\eta_0} & = & 0,
\label{eq:mostgen_backscat2}
\\
\pm\alpha_\lt{ee,A} \mp \frac{\alpha_\lt{mm,A}}{\eta_0^2}+ \frac{\alpha_\lt{em,S}}{\eta_0} + \frac{\alpha_\lt{me,S}}{\eta_0} & = & 0.
\label{eq:mostgen_backscat1}
\end{eqnarray}
Here, the top sign corresponds to the incident-wave propagation along $+z$-direction and the bottom sign along $-z$-direction.
It can be seen that the symmetric parts of the cross-coupling polarizabilities are connected to the antisymmetric parts of the electric and magnetic polarizabilities and vice versa. This is easily understood by looking at the basic equations for the dipole moments (\ref{eq:psimple}) and (\ref{eq:msimple}) as the electric and magnetic fields are orthogonal to each other in a plane wave and the antisymmetric dyadic $\overline{\overline{J}}_\lt{t}$ corresponds to a 90$^\circ$ counterclockwise rotation around the $z$-axis.

In the simplest special case  of isotropic reciprocal particles the
above relations reduce to
$\mp\alpha_\lt{ee,S}\pm\frac{\alpha_\lt{mm,S}}{\eta_0^2}=0$, which
simply means that the electric and magnetic polarizabilities should
be balanced in order to ensure the Huygens' relation between the
induced moments. For bi-anisotropic reciprocal particles, we see
from the first relation that an imbalance between these two
polarizabilities can be compensated by properly choosing the omega
coupling coefficient $\alpha_\lt{em,A}$ (or $\alpha_\lt{me,A}$ as
we have $\alpha_\lt{em,A}=\alpha_\lt{me,A}$ for omega particles).
Basically, this tells us that the required balance between the
induced electric and magnetic moments can be be achieved using the
omega coupling effect instead or in addition to the magnetization
induced by incident magnetic fields. This may be important and
beneficial for applications, because the omega coupling effect is a
first-order spatial dispersion effect, which is usually much
stronger than the second-order magnetization effect. The second
relation tells us that for reciprocal particles the zero
backscattering property is independent of the chirality parameter
$\alpha_\lt{em,S}$ as we have $\alpha_\lt{ee,A}=\alpha_\lt{mm,A}=0$
and $\alpha_\lt{em,S}=-\alpha_\lt{me,S}$ by definition.

Nonreciprocal particles can have non-zero antisymmetric parts of
electric and magnetic polarizabilities (e.g., magnetized ferrite
spheres). For particles without bi-anisotropic coupling, the second
relation tells that if the antisymmetric parts do not satisfy
$\pm\alpha_\lt{ee,A} \mp \frac{\alpha_\lt{mm,A}}{\eta_0^2}=0$, zero
backscattering cannot be achieved. In particular, this is the case
when only one of the polarizabilities exhibits nonreciprocity, but
the other one is purely symmetric (as in the same example of a
ferrite sphere). An important conclusion from this relation is that
backscattering due to nonreciprocity in electric and/or magnetic
response can be compensated by introducing magneto-electric coupling of the Tellegen type ($\alpha_\lt{em,S}=\alpha_\lt{me,S}$).

\subsubsection{Zero Forward Scattering}

Similarly, the conditions for zero forward scattering can be
derived based on (\ref{eq:zero_FS}). The condition for zero forward  scattering with arbitrary
incidence direction reads
\begin{eqnarray}
\lefteqn{\alpha_\lt{ee,S}(\unit \times \unitdyad_\lt{t})+\alpha_\lt{ee,A}(\unit \times \Jdyad_\lt{t})} \nonumber \\
& & \ \ \ +\frac{\alpha_\lt{em,S}}{\eta_0}(\unit \times \unitdyad_\lt{t} \times \unit) + \frac{\alpha_\lt{em,A}}{\eta_0}(\unit \times \Jdyad_\lt{t} \times \unit) \nonumber \\
& & \ \ \ +\frac{\alpha_\lt{me,S}}{\eta_0} \unitdyad_\lt{t} + \frac{\alpha_\lt{me,A}}{\eta_0} \Jdyad_\lt{t} \nonumber \\
& & \ \ \ +\frac{\alpha_\lt{mm,S}}{\eta_0^2}(\unitdyad_\lt{t} \times \unit) + \frac{\alpha_\lt{mm,A}}{\eta_0^2}(\Jdyad_\lt{t} \times \unit) = 0.
\label{eq:zerofwdscat_gen}
\end{eqnarray}
Again, by plugging a general directional vector $\unit =\sin{\theta}\cos{\phi}\unitx+\sin{\theta}\sin{\phi}\unity+\cos{\theta}\unitz=A\unitx+B\unity+C\unitz$ into (\ref{eq:zerofwdscat_gen}), conditions can be derived for zero forward scattering with non-axial incidence:
\begin{equation}
\alpha_\lt{em,S} = \alpha_\lt{me,S} = \alpha_\lt{ee,A} = \alpha_\lt{mm,A} = 0,
\label{eq:oblique_inc3}
\end{equation}
\begin{equation}
\alpha_\lt{ee,S} = \frac{\alpha_\lt{em,A}}{\eta} C = -\frac{\alpha_\lt{me,A}}{\eta} \frac{1}{C} = \frac{\alpha_\lt{mm,S}}{\eta^2}. \hspace{0.8cm}
\label{eq:oblique_inc4}
\end{equation}
As with zero backscattering, only one angle of incidence can be covered with given polarizabilities and this solution is only valid for an ideally uniaxial particle.

For the axial incidence, we get two scalar equations:
\begin{eqnarray}
\pm\alpha_\lt{ee,S}\pm\frac{\alpha_\lt{mm,S}}{\eta_0^2} - \frac{\alpha_\lt{em,A}}{\eta_0} + \frac{\alpha_\lt{me,A}}{\eta_0} & = & 0, \hspace{0.5cm}
\label{eq:mostgen_fwdscat2}
\\
\mp\alpha_\lt{ee,A} \mp \frac{\alpha_\lt{mm,A}}{\eta_0^2}- \frac{\alpha_\lt{em,S}}{\eta_0} + \frac{\alpha_\lt{me,S}}{\eta_0} & = & 0,
\label{eq:mostgen_fwdscat1}
\end{eqnarray}
where, again, the top sign corresponds to the incident-wave propagation along $+z$-direction and the bottom sign along $-z$-direction. Notably, these conditions greatly resemble the conditions for zero backscattering with the only difference being the signs of some of the components.

In this case, an imbalance between the electric and magnetic
polarizabilities (the first relation) can be compensated by
magneto-electric coupling coefficient of a moving-particle type ($\alpha_\lt{em,A}=-\alpha_\lt{me,A}$).
Forward scattering due to antisymmetric (nonreciprocal) parts of
electric and magnetic polarizabilities can be compensated by making the particle chiral (i.e., introducing magneto-electric coupling of the type $\alpha_\lt{em,S}=-\alpha_\lt{me,S}$) and properly choosing the value of the chirality parameter (the
second relation). In other words, nonreciprocal Faraday rotation in transmission can be compensated by reciprocal rotation due to the chirality of the particle.
\subsubsection{Zero Total Scattering}
But what if we demand that both  forward and backward scattering
amplitudes are zero? By demanding all the equations
(\ref{eq:mostgen_backscat2}), (\ref{eq:mostgen_backscat1}), (\ref{eq:mostgen_fwdscat2}),
and (\ref{eq:mostgen_fwdscat1}) to be true, we get a set of four
equations for the eight unknown polarizability components:
\begin{eqnarray}
\alpha_\lt{ee,S} &=& \pm\frac{\alpha_\lt{em,A}}{\eta_0},
\label{eq:mostgen_totscat1}
\\
\alpha_\lt{mm,S} &=& \mp\eta_0 \alpha_\lt{me,A},
\label{eq:mostgen_totscat3}
\\
\alpha_\lt{ee,A} &=& \mp\frac{\alpha_\lt{em,S}}{\eta_0},
\label{eq:mostgen_totscat2}
\\
\alpha_\lt{mm,A} &=& \pm\eta_0 \alpha_\lt{me,S},
\label{eq:mostgen_totscat4}
\end{eqnarray}
where the top sign corresponds to the incident-wave propagation
along $+z$-direction and the bottom sign along $-z$-direction. By
plugging (\ref{eq:mostgen_totscat1})--(\ref{eq:mostgen_totscat4})
into (\ref{eq:zero_sca_Fsca3}), it can be seen that these conditions
hold not only for zero scattering in the forward and backward
directions but actually for zero scattering in all directions.
Actually, if
(\ref{eq:mostgen_totscat1})--(\ref{eq:mostgen_totscat4}) are
satisfied, the induced dipole moments are equal to zero. This means
that such particles are not excited by plane waves propagating
axially with the particle axis in one direction (i.e., no electric
nor magnetic dipole moments induced), but they show non-zero total
scattering when the incident wave is propagating in the opposite
direction. In fact, it can be seen that for the opposite axial propagation direction the zero backscattering condition is satisfied as long as we have $\adyad_\lt{em} = \adyad_\lt{me}$. On the other hand, if we have $\adyad_\lt{em} = -\adyad_\lt{me}$, the zero forward scattering condition is satisfied instead.

As zero forward scattering property is governed by the optical
theorem, the topic merits further investigation. Here, we will study
passivity of zero forward scattering particles and clarify the
physical meaning of the derived zero forward scattering conditions.
The passivity of a particle can be studied by looking at the power
spent by the incident wave for exciting the particle, which is given
by \cite{Yatsenko}
\begin{equation}
P=\frac{1}{2} \Re[{(j \omega \mathbf{p})^* \cdot \mathbf{E}_\lt{inc} + j \omega \mathbf{m} \cdot \mathbf{H}_\lt{inc}^*}].
\label{eq:absorbedP}
\end{equation}%
If the particle absorbs power (some of which may be re-radiated by
the particle), we have $P > 0$ and if it generates power (i.e., is
active), we have $P < 0$. If we have $P = 0$, there is no power
exchange between the incident wave and the particle. By plugging the
dipole moments generated by a general bi-anisotropic particle when
the zero forward scattering conditions (\ref{eq:mostgen_fwdscat2}) and (\ref{eq:mostgen_fwdscat1}) are satisfied into
(\ref{eq:absorbedP}), passivity of the particle can be analyzed.

Firstly, by using the definitions
(\ref{eq:aee_gen})--(\ref{eq:ame_gen}), equation
(\ref{eq:absorbedP}) can be written as
\begin{equation}
P=\frac{1}{2} \omega |E_\lt{inc}|^2 \Im \left[\alpha^*_\lt{ee,S}\mp\frac{\alpha_\lt{em,A}^*}{\eta_0}-\frac{\alpha_\lt{mm,S}}{\eta_0^2} \mp \frac{\alpha_\lt{me,A}}{\eta_0} \right],
\label{eq:absorbedP_gen1}
\end{equation}%
where the top sign corresponds to propagation along $+z$-direction
and the bottom sign along $-z$-direction. If we now enforce the zero
forward scattering conditions (\ref{eq:mostgen_fwdscat2}) and
(\ref{eq:mostgen_fwdscat1}), we see that the power absorbed by the
particle is identically zero for any values of the polarizabilities.
Note that the induced dipole moments are not zero, except if also
the backward scattering is zero. Therefore, there is seemingly no
power interaction between the incident wave and the particle. As the
particle takes no power from the incident field, it cannot
re-radiate any power either. Still, the particle is not necessarily
invisible. This can be seen by looking at the power radiated by the
particle \cite{Yatsenko}
\begin{equation}
P=\frac{\omega^4}{12 \pi c} (\mu_0 \mathbf{p}^* \cdot \mathbf{p} + \epsilon_0 \mathbf{m}^* \cdot \mathbf{m}).
\label{eq:radiatedP}
\end{equation}%
For a passive particle, this power should be equal to the power
spent by the incident power for exciting the particle. However, the
radiated power is zero only if the induced dipole moments are both
zero and this is only true if the zero total scattering conditions
are satisfied. Therefore, in order to have exactly zero forward
scattering with non-zero total scattering, the particle has to be
active. As discussed in \cite{AluEng,NV_JOSA,NV_nano}, this does not exclude
possibilities for achieving greatly reduced (but still not exactly
zero) forward scattering with non-zero backward and total
scattering. Basically, this is achieved when the real part of
(\ref{eq:mostgen_fwdscat2}) is zero and the imaginary part of
(\ref{eq:mostgen_fwdscat1}) is zero. Since the optical theorem is
only valid for passive particles, the results presented in this
paper are not in conflict with the optical theorem.

In the following, we specify the general results for all fundamental
classes of bi-anisotropic particles.

\subsection{Uniaxial Omega Particle and Uniaxial Moving Particle}

For a uniaxial omega particle along the $z$-axis, the polarizabilities have the form
\begin{equation}
\left\{\begin{array}{l}
\displaystyle
\aee=\alpha_\lt{ee}\overline{\overline{I}}_\lt{t}
\vspace*{.2cm}\\\displaystyle
\amm=\alpha_\lt{mm}\overline{\overline{I}}_\lt{t}
\vspace*{.2cm}\\\displaystyle
\aem=\alpha_\lt{em}\overline{\overline{J}}_\lt{t}
\vspace*{.2cm}\\\displaystyle
\ame=\alpha_\lt{me}\overline{\overline{J}}_\lt{t},
\end{array}\right. \label{eq:omegamoving}
\end{equation}
that is, only the symmetric parts of $\aee$ and $\amm$ and antisymmetric parts of $\aem$ and $\ame$ are non-zero. The omega particle is reciprocal, i.e., we have
\begin{equation}
\aem=\ame=\alpha \overline{\overline{J}}_\lt{t}.
\label{eq:omegareciprocal}
\end{equation}
Equations (\ref{eq:omegamoving}) hold also for a uniaxial ``moving'' particle (with the axis oriented along the $z$-axis), and it is, therefore, convenient to study these two classes together. However, as a uniaxial ``moving'' particle is nonreciprocal, we must enforce, instead of (\ref{eq:omegareciprocal}), the condition
\begin{equation}
\aem=-\ame=\alpha \overline{\overline{J}}_\lt{t}.
\label{eq:movingnonreciprocal}
\end{equation}
``Moving'' is set here inside quotation marks as the ``moving'' particle is not a particle in motion, but a nonreciprocal particle at rest with the polarizabilities in this form. These quotation marks are dropped in the rest of the paper for simplicity of notations.

\subsubsection{Zero Backscattering}

Let us first consider backscattering from a single uniaxial  omega
or moving particle. As (\ref{eq:oblique_inc1}) and
(\ref{eq:oblique_inc2}) are not satisfied, zero backscattering is
not possible for oblique incidence for either particle. For the
axial incidence ($\unit = \pm \unitz$), the zero backscattering
condition (\ref{eq:mostgen_backscat2}) simplifies to
\begin{equation}
-\alpha_\lt{ee}\pm\frac{2}{\eta_0}\alpha +\frac{1}{\eta_0^2}\alpha_\lt{mm}=0
\label{eq:omega_zerobackscat}
\end{equation}
for an omega particle and to
\begin{equation}
\alpha_\lt{ee}=\frac{1}{\eta_0^2}\alpha_\lt{mm}, \quad \mbox{and arbitrary} \, \, \alpha
\label{eq:moving_zerobackscat}
\end{equation}
for a moving particle. Here,  the $\pm$ signs correspond to the
opposite directions of the incident plane wave ($\unit = \pm
\unitz$). The other zero backscattering condition
(\ref{eq:mostgen_backscat1}) vanishes for both particles. As the
scattering losses must be included according to \r{5}--\r{8}, all
the polarizabilities can be complex quantities even for a particle
without absorption losses. Therefore, both equations can, at least
in principle, be satisfied even if the absorption losses are
excluded. However, if scattering and absorption losses are both
excluded, (\ref{eq:omega_zerobackscat}) cannot be satisfied for
$\alpha\neq 0$. It should be noted that the condition for moving
particles holds also if the propagation direction is reversed
despite of the inherent nonreciprocity of the particle. Also, as the
self-duality conditions $\amm =  \eta_0^2 \aee$ and $\aem  =  -\ame$
as well as the $\pi/2$ rotational symmetry condition with the
axial incidence are both satisfied by definition for a moving
particle, we can conclude that the general conditions for zero
backscattering derived in \cite{Lindell2009} hold for the moving
particle case. On the other hand, by definition these conditions are
not satisfied for an omega particle, as we have $\adyad_\lt{em} =
\adyad_\lt{me}$, not $\adyad_\lt{em} = -\adyad_\lt{me}$. However, in
\cite{Lindell2009} in deriving the self-duality condition the
particles were assumed to be lossless. If the losses were to be
included in the derivation, our results would agree with the earlier
results also for the omega particle. This makes sense as satisfying
(\ref{eq:omega_zerobackscat}) and (\ref{eq:moving_zerobackscat}) for
the corresponding particles results in the same induced electric and
magnetic dipole moments in both cases.

\begin{figure}[htb]
  \centering
  \includegraphics[trim = 0 0.0cm 0cm 0.0cm, clip = false, width=0.35\textwidth]{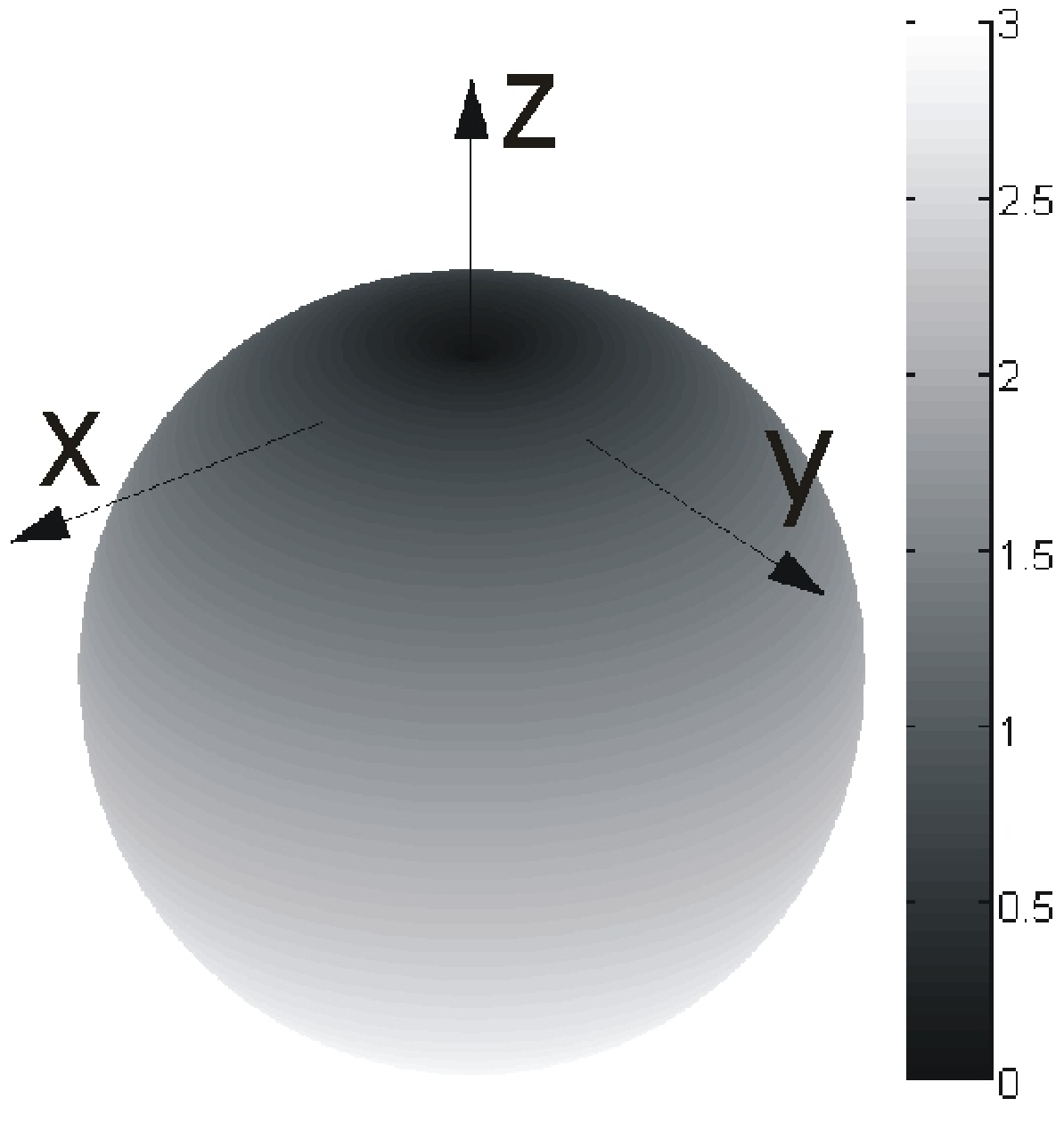} 
  \caption{Scattering directivity pattern (absolute values) for an incident plane wave propagating along the $- z$-direction exciting a moving particle with the polarizabilities satisfying $\alpha_\lt{ee}=\frac{1}{\eta_0^2}\alpha_\lt{mm}$ (or an omega particle with ${-\alpha_\lt{ee}-\frac{2}{\eta_0}\alpha+\frac{1}{\eta_0^2}\alpha_\lt{mm}=0}$).}
  \label{fig:1}
\end{figure}

Next, we study the scattering directivity pattern of particles which satisfy the zero-scattering conditions. To do that, we substitute the dipole moments (\ref{eq:psimple}) and (\ref{eq:msimple}) into (\ref{eq:Fsca}) and use 
(\ref{eq:H_from_E}) to eliminate the incident magnetic field.
If we enforce the condition (\ref{eq:omega_zerobackscat}) for an omega particle or (\ref{eq:moving_zerobackscat}) for a moving particle, the radiation vector for the axial incidence with an arbitrary radiation direction $\vt{u}_\lt{r}$ reduces in both cases to
\begin{equation}
\vt{F}_\lt{scat} = (\alpha_\lt{ee}\mp\alpha/\eta_0)[(\unitr \times \unitdyad_\lt{t})\pm\overline{\overline{J}}_\lt{t}] \cdot \vt{E}_\lt{inc}
\label{eq:radvector1}
\end{equation}
where the top sign corresponds to the incident wave propagation
direction $+\unitz$ and  the bottom sign to $-\unitz$. From this
result, we can see that the scattered wave always has linear
polarization for a linearly polarized incident wave. This means that
the axial ratio of the scattered wave, defined here as the ratio of
the lengths of the minor and major polarization axes, is zero.
However, the polarization plane of the wave is rotated. The
corresponding scattering directivity pattern is plotted in Fig.~\ref{fig:1} for
a plane wave propagating along $-z$-direction. It can be observed
that the maximum directivity is in the forward direction and has the
absolute value of 3. For the opposite incidence direction, the
scattering pattern relative to the incidence direction is the same,
i.e., the pattern is simply flipped with respect to the incidence
direction.

\subsubsection{Zero Forward Scattering}

Similarly, we can derive conditions for zero forward scattering for
these two classes of uniaxial particles. Again, zero forward
scattering at oblique incidence is not possible as
(\ref{eq:oblique_inc3}) and (\ref{eq:oblique_inc4}) are not
satisfied. For the axial incidence with $\unit = \unitr = \pm
\unitz$, the zero forward scattering condition
(\ref{eq:mostgen_fwdscat2}) simplifies to
\begin{equation}
\alpha_\lt{ee}=-\frac{1}{\eta_0^2}\alpha_\lt{mm} \quad \mbox{and arbitrary} \, \, \alpha
\label{eq:omega_zerofwdscat}
\end{equation}
for an omega particle and
\begin{equation}
-\alpha_\lt{ee}\pm\frac{2}{\eta_0}\alpha - \frac{1}{\eta_0^2}\alpha_\lt{mm}=0
\label{eq:moving_zerofwdscat}
\end{equation}
for a moving particle. The other zero forward  scattering condition
(\ref{eq:mostgen_fwdscat1}) vanishes for both particles. Here we should remember that the polarizabilities in 
(\ref{eq:omega_zerofwdscat}) and (\ref{eq:moving_zerofwdscat}) are complex numbers due to scattering losses. Thus,
the
condition for an omega particle cannot be realized, at least with a
conventional passive uniaxial omega particle (realized, for example,
as two orthogonal $\Omega$-shaped pieces of metal wire), because the
imaginary parts of both polarizabilities have the same sign. 
Interestingly, the relation between the electric and magnetic
polarizabilities (\ref{eq:omega_zerofwdscat}) is the same condition
that was derived for an electrically small isotropic
magnetodielectric spherical particle in \cite{Kerker}. In that
paper, the condition is given as $\epsilon_r =
(4-\mu_r)/(2\mu_r+1)$. This reduces to our condition as the
normalized electric and magnetic polarizabilities are
$\alpha_\lt{ee} =(\epsilon_r-1)/(\epsilon_r+2)$ and $\alpha_\lt{mm}
=(\mu_r-1)/(\mu_r+2)$ for a small magnetodielectric spherical
particle. Our result shows that this condition holds also for
bi-anisotropic particles with omega coupling for arbitrary values of
the coupling coefficient. As discussed above,  the zero forward
scattering conditions derived here can be satisfied only
approximately with passive particles, in harmony with the optical
theorem. For the special case of simple isotropic particles
this issue was discussed in \cite{AluEng,NV_JOSA}.

Noticeably for a moving particle, we get different solutions for different incidence directions, similar to the omega particle case for zero backscattering. Again, the induced dipole moments are the same for both particles when the corresponding zero forward scattering conditions are satisfied. The dipole moments are non-zero meaning that there still can exist scattering in other directions.  

If we enforce the zero forward scattering conditions (\ref{eq:omega_zerofwdscat}) for an omega particle or (\ref{eq:moving_zerofwdscat}) for a moving particle, the radiation vector for the axial incidence ($\unit = \pm \unitz$) and an arbitrary scattering direction $\unitr$ has the form
\begin{equation}
\vt{F}_\lt{scat} = (\alpha_\lt{ee}\mp\alpha/\eta_0)[(\unitr \times \unitdyad_\lt{t})\mp\overline{\overline{J}}_\lt{t}] \cdot \vt{E}_\lt{inc}.
\label{eq:radvector2}
\end{equation}
It can be observed that this radiation vector has a form similar to (\ref{eq:radvector1}) with only the sign of the $\overline{\overline{J}}_\lt{t}$--term changed. Therefore, the scattering directivity pattern is also the same as the one shown in Fig.~\ref{fig:1} albeit flipped so that there is now no scattering in the forward direction. Also, the polarization of the scattered field is linear as before.
\subsubsection{Zero Total Scattering}
In the case of the omega particle, the zero total scattering conditions (\ref{eq:mostgen_totscat1})--(\ref{eq:mostgen_totscat4}) simplify to
\begin{equation}
\alpha_\lt{ee}=\pm\frac{1}{\eta_0} \alpha=-\frac{1}{\eta_0^2} \alpha_\lt{mm}
\label{eq:omega_zerototal}
\end{equation}
which, again, cannot be fulfilled with a conventional passive omega particle,  since the imaginary parts of $\alpha_\lt{ee}$ and $\alpha_\lt{mm}$ of passive particles have the same sign and they cannot be zero if particles receive and scatter power.
In the case of a moving particle, we get
\begin{equation}
\alpha_\lt{ee}=\pm \frac{1}{\eta_0} \alpha = \frac{1}{\eta_0^2} \alpha_\lt{mm}.
\label{eq:moving_zerototal}
\end{equation}
which can be in principle satisfied, giving zero induced dipole
moments. It is interesting to note that these conditions are the
same as the conditions for the ``optimal'' bi-anisotropic particles
whose energy in a given external field is maximized or minimized \cite{optimal}. Also, if we satisfy zero total scattering conditions
(\ref{eq:omega_zerototal}) or (\ref{eq:moving_zerototal}) for one
axial propagation direction, a plane wave propagating in the
opposite direction scatters from the particle, but the forward
scattering amplitude in the case of an omega particle or the
backscattering amplitude in the case of a moving particle is zero,
because the zero forward scattering condition,
(\ref{eq:omega_zerofwdscat}), or the zero backscattering condition,
(\ref{eq:moving_zerobackscat}), is satisfied for both propagation
directions. As it was shown earlier that zero forward scattering
with non-zero total scattering requires an active particle,  this
also implies that an invisible omega particle has to be active
whereas an invisible moving particle does not have to be.

\subsection{Uniaxial Chiral Particle and Uniaxial Tellegen Particle}

Next, we consider uniaxial chiral particles and uniaxial Tellegen
particles oriented along the $z$-axis, with the polarizabilities of
the form
\begin{equation}
\left\{\begin{array}{l}
\displaystyle
\aee=\alpha_\lt{ee}\overline{\overline{I}}_\lt{t}
\vspace*{.2cm}\\\displaystyle
\amm=\alpha_\lt{mm}\overline{\overline{I}}_\lt{t}
\vspace*{.2cm}\\\displaystyle
\aem=\alpha_\lt{em}\overline{\overline{I}}_\lt{t}
\vspace*{.2cm}\\\displaystyle
\ame=\alpha_\lt{me}\overline{\overline{I}}_\lt{t},
\end{array}\right. \label{eq:chiralTellegen1}
\end{equation}
that is, only the symmetric parts of the general  polarizability
dyadics (\ref{eq:aee_gen})--(\ref{eq:ame_gen}) are non-zero.
Furthermore, we have
\begin{equation}
\aem=\pm \ame=\alpha \overline{\overline{I}}_\lt{t},
\label{eq:chiralTellegen2}
\end{equation}
where ($+$) corresponds to a Tellegen particle and ($-$) to a chiral particle.

\subsubsection{Zero Backscattering}

Based on (\ref{eq:mostgen_backscat2}) and (\ref{eq:mostgen_backscat1}), the zero backscattering condition for a chiral particle assuming axial incidence ($\unit = \pm \unitz$) reads:
\begin{equation}
\alpha_\lt{ee}=\frac{1}{\eta_0^2}\alpha_\lt{mm} \quad \mbox{and arbitrary} \, \, \alpha .
\label{chiral_zerobackscat}
\end{equation}
As the condition $\adyad_\lt{em} = -\adyad_\lt{me}$ as well  as the
rotational symmetry condition are satisfied by definition for the
chiral particle, we can conclude that the result is in agreement
with the earlier results of \cite{Karilainen_chiral,Lindell2009}.
Also, as (\ref{eq:oblique_inc1}) and (\ref{eq:oblique_inc2}) are not
satisfied, zero backscattering is only possible for the axial
incidence. The radiation vector for the axial  incidence ($\unit =
\pm \unitz$) when the zero backscattering condition is satisfied has
the form
\begin{equation}
\vt{F}_\lt{scat} = [\alpha_\lt{ee}((\unitr \times \unitdyad_\lt{t}) \pm \overline{\overline{J}}_\lt{t}) + \frac{\alpha}{\eta_0}(-\unitdyad_\lt{t} \pm (\unitr \times \overline{\overline{J}}_\lt{t})] \cdot \vt{E}_\lt{inc}.
\label{eq:radvector3}
\end{equation}
In this case, the scattered wave excited by a linearly polarized plane wave is, in general case, elliptically polarized. The axial ratio of the scattered wave depends on the values of polarizabilities $\alpha_\lt{ee}$ and $\alpha$. Because the zero backscattering condition does not include the field coupling coefficient $\alpha$, the axial ratio of the scattered field can be tuned by choosing the value of the chirality parameter.

As (\ref{eq:mostgen_backscat1}) has no solution for a Tellegen
particle (except with $\alpha = 0$), zero backscattering from a
Tellegen particle excited by a linearly polarized plane wave is not
possible.

\subsubsection{Zero Forward Scattering}
Similarly, we can analyze conditions for zero forward scattering. In
this case, equations
(\ref{eq:mostgen_fwdscat2})--(\ref{eq:mostgen_fwdscat1}) reduce for
a Tellegen particle to
\begin{equation}
\alpha_\lt{ee}=-\frac{1}{\eta_0^2}\alpha_\lt{mm} \quad \mbox{and arbitrary} \, \, \alpha ,
\label{eq:Tellegen_zerofwdscat}
\end{equation}
which is the same condition  that we got for omega particles. Again, as (\ref{eq:oblique_inc3}) and (\ref{eq:oblique_inc4}) are not satisfied
for a Tellegen particle, zero forward scattering is only possible for the axial incidence. 
In the Tellegen particle case, the radiation vector for the axial  incidence ($\unit = \pm \unitz$) when the zero forward scattering condition is satisfied has the form
\begin{equation}
\vt{F}_\lt{scat} = [\alpha_\lt{ee}((\unitr \times \unitdyad_\lt{t}) \mp \overline{\overline{J}}_\lt{t}) + \frac{\alpha}{\eta_0}(\unitdyad_\lt{t} \pm (\unitr \times \overline{\overline{J}}_\lt{t})] \cdot \vt{E}_\lt{inc}.
\label{eq:radvector4}
\end{equation}
Once more, the scattering directivity pattern is the same as shown in
Fig.~\ref{fig:1} (in this case corresponding to a wave propagating
along the $+z$-direction) and the scattered wave is elliptically
polarized. Also in this case we note that the polarization can be
tuned by adjusting the field coupling coefficient, because
(\ref{eq:Tellegen_zerofwdscat}) holds independently of the value of
$\alpha$. As (\ref{eq:mostgen_fwdscat1}) has no solution for a
chiral particle (except $\alpha=0$), zero forward scattering is
not possible in this case.

\begin{table*}[!t]
\caption{Conditions for zero backscattering, zero forward scattering, and zero
total scattering (if available), the axial ratio of the scattered
wave when the corresponding condition is met (axial incidence) and the passivity/activity of the required particle. The
top sign corresponds to the incident wave propagating along the
$+z$-direction and the bottom sign along $-z$-direction. The
incident plane wave is linearly polarized. $AR$ indicates the values
of the axial ratio of the scattered field, where $AR=0$ corresponds
to linear polarization.} \label{T:2}
\begin{tabular}{|c|p{36mm}|p{41.9mm}|p{36mm}|p{41.9mm}|}
\hline
\rowcolor[gray]{.9}
                & Chiral particle   & Omega particle    & Tellegen particle & Moving particle \\

\hline
\begin{turn}{270} backscattering \hspace{2pt} \end{turn}%
\begin{turn}{270} Zero \hspace{2pt} \end{turn}%
\cellcolor[gray]{.9}

& \vspace{0.5mm} $\alpha_\lt{ee} = 1/\eta_0^2 \alpha_\lt{mm}$ \newline \newline $-1 \leq AR \leq 1$ \newline \newline passive
& \vspace{0.5mm} $-\alpha_\lt{ee} \pm\frac{2}{\eta_0}\alpha +\frac{1}{\eta_0^2}\alpha_\lt{mm}=0$  
\newline \newline $AR = 0$ \newline \newline passive 
& 
\cellcolor[gray]{.3}
& \vspace{0.5mm} $ \alpha_\lt{ee} = 1/\eta_0^2 \alpha_\lt{mm}$ \newline \newline $AR = 0$ \newline \newline passive  \\
\hline
\begin{turn}{270}scattering \ \ \ \ \ \ \hspace{2pt} \end{turn}%
\begin{turn}{270}Zero forward \hspace{2pt} \end{turn}%
\cellcolor[gray]{.9}

& 
\cellcolor[gray]{.3}
& \vspace{0.5mm} $\alpha_\lt{ee} = -1/\eta_0^2 \alpha_\lt{mm}$ \newline \newline $AR = 0$  \newline \newline active
& \vspace{0.5mm} $\alpha_\lt{ee}=-1/\eta_0^2\alpha_\lt{mm}$ \newline \newline $-1 \leq AR \leq 1$ \newline \newline active
& \vspace{0.5mm} $-\alpha_\lt{ee} \pm \frac{2}{\eta_0}\alpha - \frac{1}{\eta_0^2}\alpha_\lt{mm}=0$ 
\newline \newline $AR = 0$  \newline \newline active \newline (unless also $\alpha_\lt{ee}=1/\eta_0^2 \alpha_\lt{mm}$) \\
\hline
\begin{turn}{270}scattering \ \ \ \ \ \ \hspace{2pt} \end{turn}%
\begin{turn}{270}Zero total \hspace{2pt} \end{turn}%
\cellcolor[gray]{.9}
& 
\cellcolor[gray]{.3}
%
& \vspace{0.5mm} $\alpha_\lt{ee} = \pm 1/\eta_0 \alpha= -1/\eta_0^2 \alpha_\lt{mm}$  
\newline \newline $AR = 0$ \newline \newline active 
&
\cellcolor[gray]{.3}
& \vspace{0.5mm} $\alpha_\lt{ee} = \pm 1/\eta_0 \alpha= 1/\eta_0^2 \alpha_\lt{mm}$ 
\newline \newline $AR = 0$ \newline \newline passive\\
\hline
\end{tabular}
\label{ta:load_values}
\end{table*}

\section{Simulation Results}

In this section, we will demonstrate how zero backscattering and zero forward scattering conditions can be realized in practice with actual particles. We will use Ansys HFSS software to simulate the scattering of an incident linearly polarized plane wave from two reciprocal particles, one corresponding to zero backscattering (chiral particle) and the other one to zero forward scattering (omega particle). The material of the particles in this model is perfect electric conductor (PEC).

\begin{figure}[htb]
  \centering
  \includegraphics[width=0.31\textwidth]{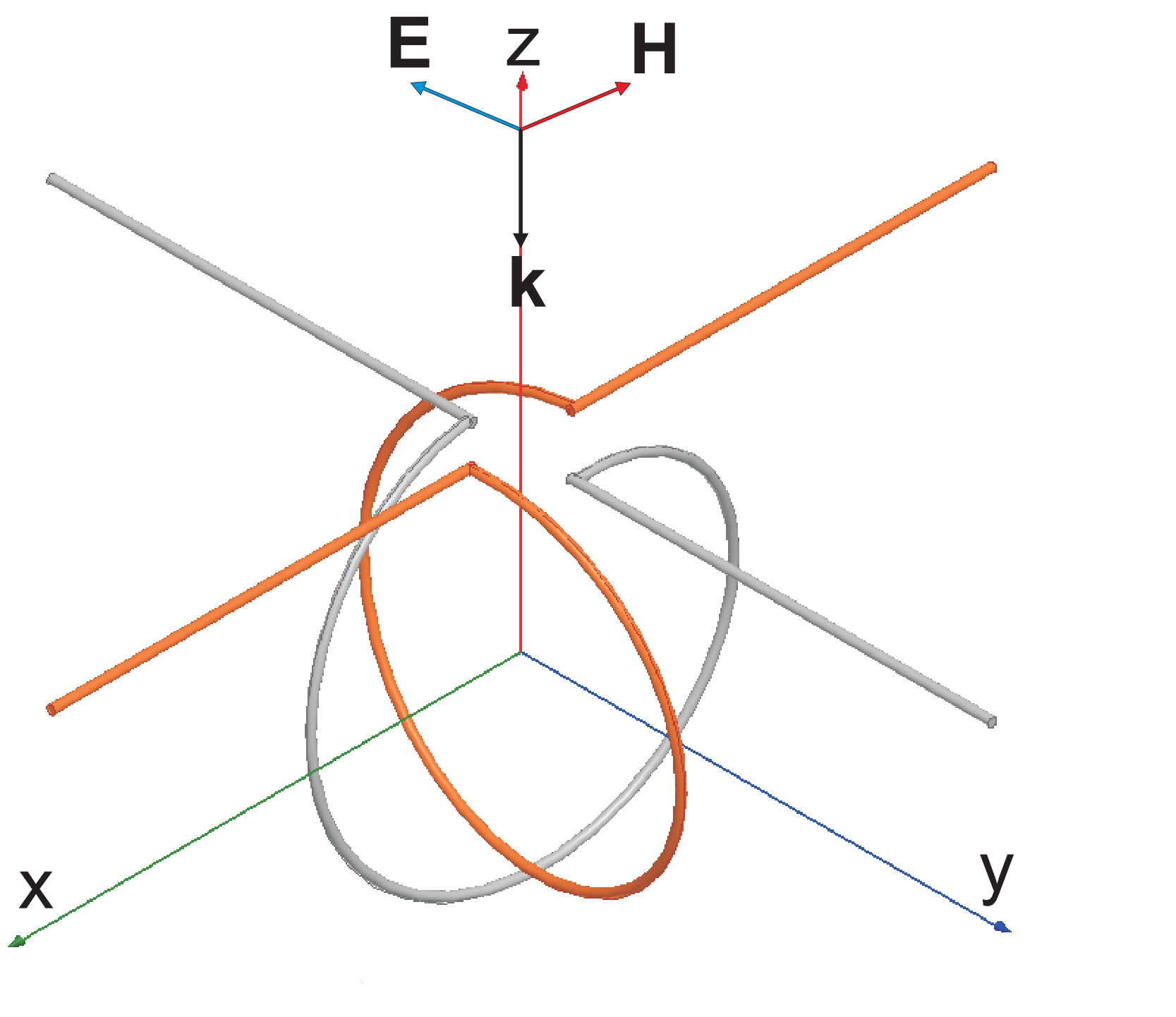} \\
  \caption{Uniaxial chiral particle showing zero backscattering at 1.8~GHz. The two elements are drawn with different colors only for clarity.}
  \label{fig:chiral}
\end{figure}

\begin{figure}[ht!]
  \centering
  \subfigure[]{\includegraphics[trim = 1cm 0cm .2cm 2.4cm,width=0.241\textwidth]{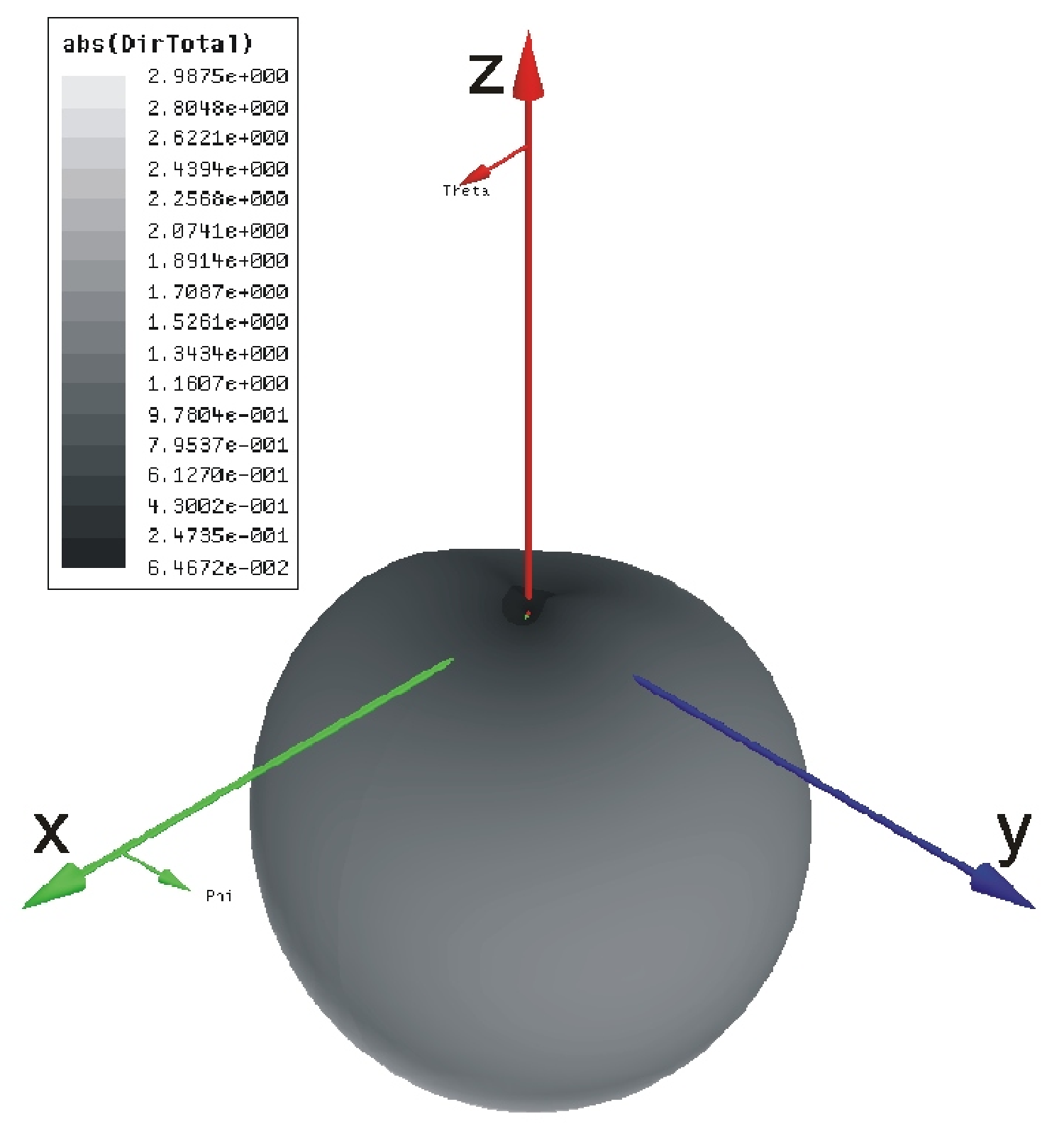}}  
  \subfigure[]{\includegraphics[trim = 1cm 0cm 1cm 2.4cm,width=0.241\textwidth]{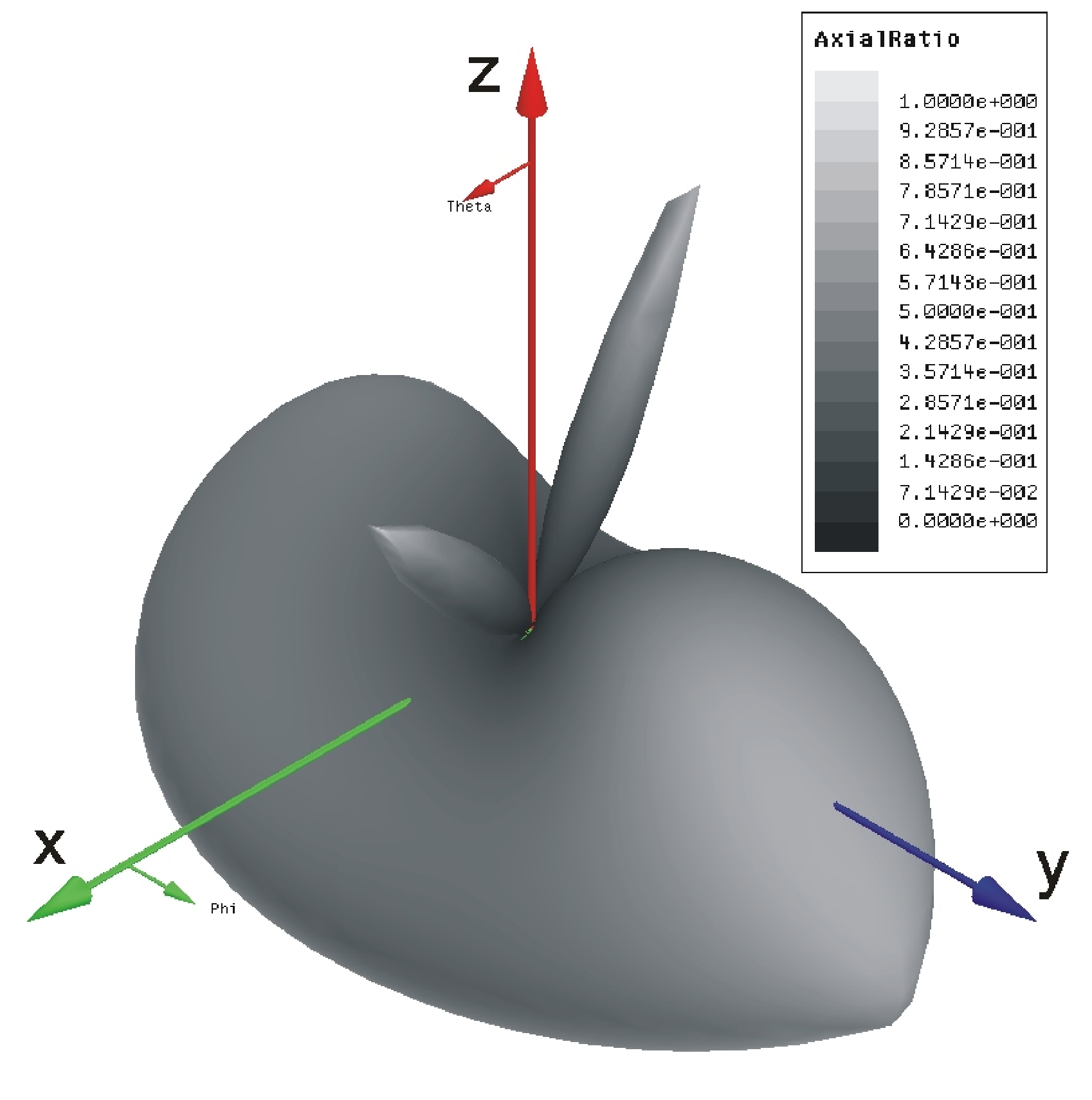}} \\
  \caption{Scattering directivity pattern (a) and axial ratio pattern (b) for the chiral particle of Fig.~\ref{fig:chiral} at 1.8~GHz (absolute values). }
  \label{fig:chiral2}
\end{figure}

It is well known that a uniaxial particle with zero backscattering can be realized as two short dipole antennas connected to orthogonally oriented small loop antennas as long as the dipole and loop dimensions are chosen so that the particle is balanced, i.e., the condition (\ref{chiral_zerobackscat}) is satisfied \cite{Karilainen_chiral,Karilainen_antenna,Lindell2009}. An example design operating at the frequency 1.8~GHz is shown in Fig.~\ref{fig:chiral}. The length of one dipole arm is 20~mm, the loop radius is 8.75~mm, the radius of the wire is 0.2~mm, and the separation between the dipole arms is 5~mm. The corresponding scattering directivity pattern is shown in Fig.~\ref{fig:chiral2}(a). Clearly, the scattering to the backward direction is minimized. Also, the shape of the scattering directivity pattern is very close to the ideal pattern of Fig.~\ref{fig:1} and the maximum directivity is close to the theoretical value of 3. The scattering directivity in the backward direction is 0.18. The corresponding axial ratio pattern (absolute values) is shown in Fig.~\ref{fig:chiral2}(b). Ideally, we should expect the axial ratio of 1 (i.e., the scattered field should have circular polarization) for lossless balanced chiral particles, but due to unwanted coupling between the two elements the axial ratio in the forward direction is only 0.77. This could be increased by, e.g., increasing the separation between the dipole arms. However, this would negatively affect the scattering directivity pattern.

\begin{figure}[htb]
  \centering
  \includegraphics[width=0.35\textwidth]{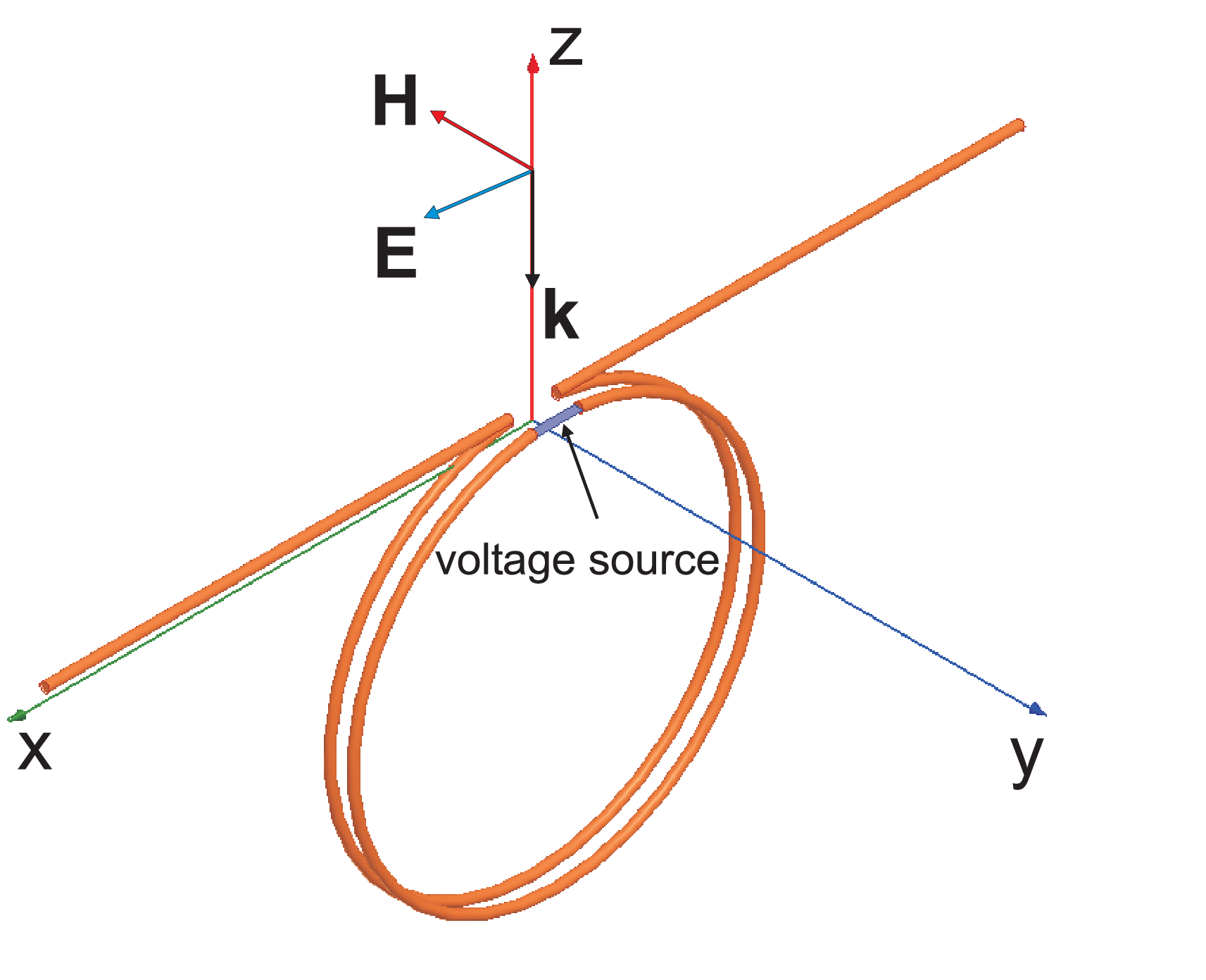} \\
  \caption{Active omega particle showing zero forward scattering at 1.8~GHz.}
  \label{fig:omega}
\end{figure}

\begin{figure}[ht!]
  \centering
  \subfigure[]{\includegraphics[trim = 1cm 0cm .4cm 0cm,width=0.241\textwidth]{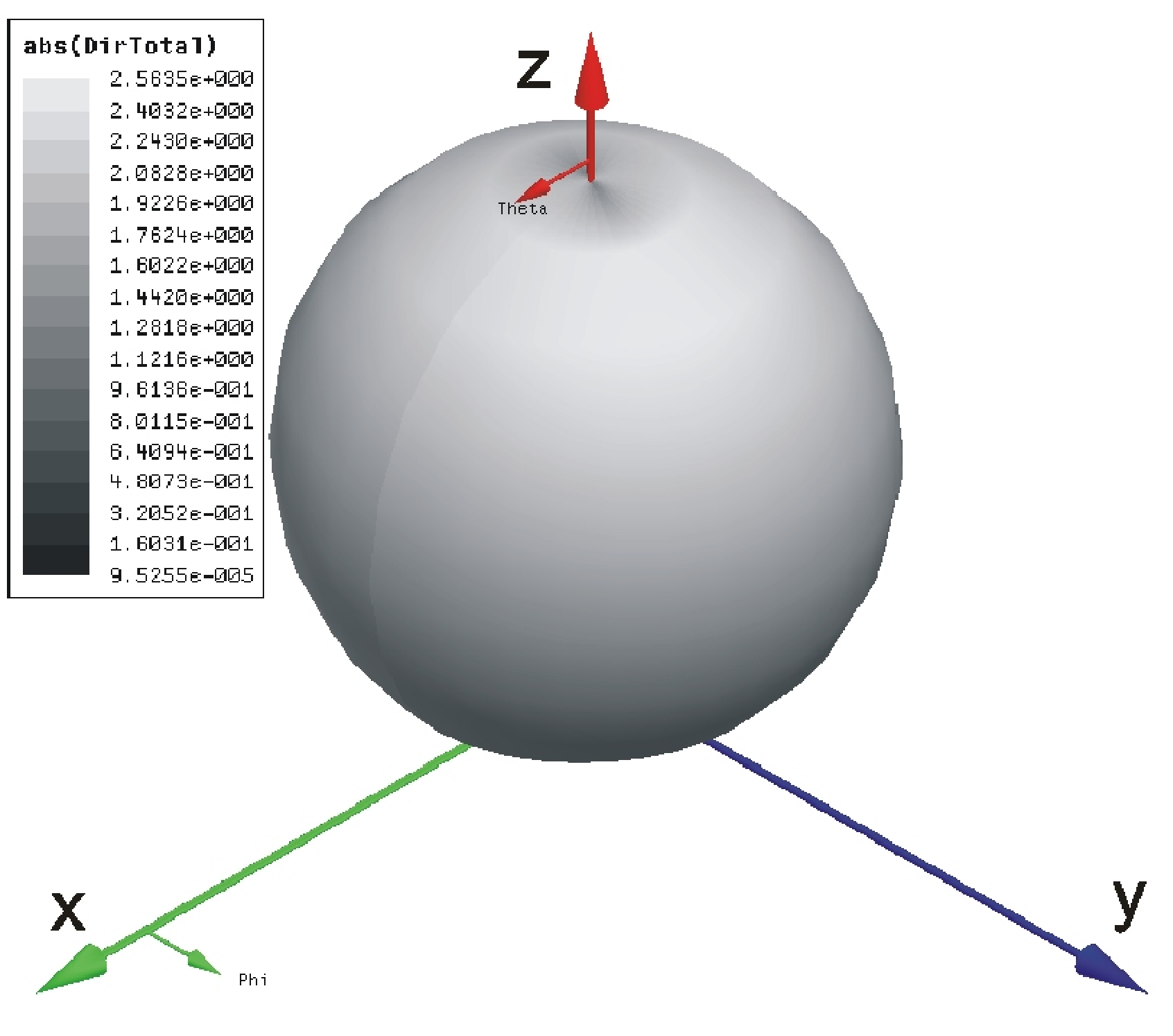}}  
  \subfigure[]{\includegraphics[trim = 1cm 0cm 1cm 0cm,width=0.241\textwidth]{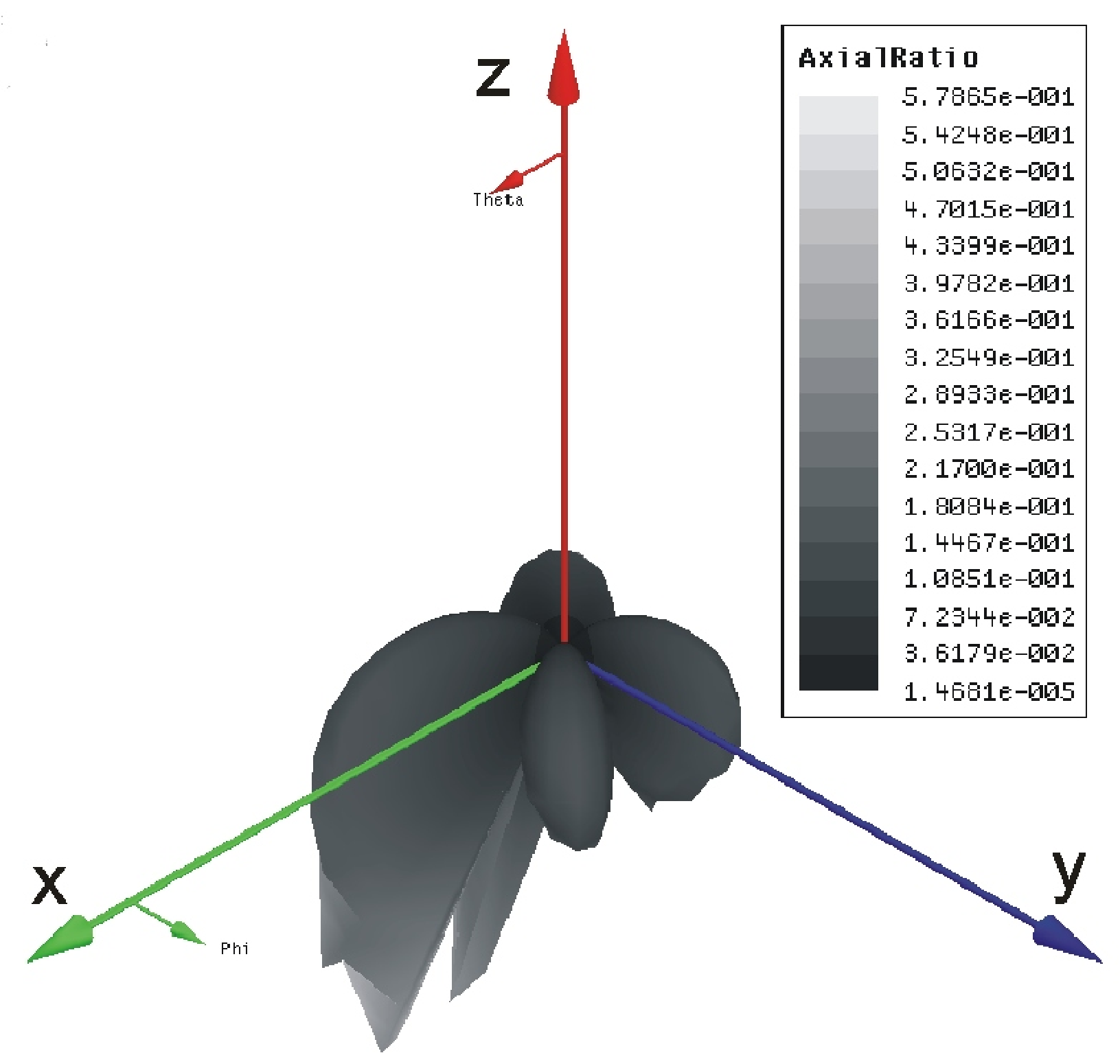}} \\
  \caption{Scattering directivity pattern (a) and axial ratio pattern (b) for the active omega particle of Fig.~\ref{fig:omega} at 1.8~GHz (absolute values).}
  \label{fig:omega2}
\end{figure}

As mentioned earlier, particles with zero forward scattering cannot be realized with passive inclusions. However, we can create a zero forward scattering particle by adding an active element to a conventional metal omega particle (in our example, a canonical omega particle consisting of two orthogonal $\Omega$-shaped metal wires). To be precise, this can be achieved by placing a voltage-driven wire loop next to an omega particle. If the external voltage applied to the loop has the opposite direction to the incident electric field and the appropriate amplitude and phase, we can effectively change the sign of the magnetic polarizability $\alpha_\lt{mm}$, thus fulfilling the required condition for zero forward scattering for an omega particle (\ref{eq:omega_zerofwdscat}). An example design for such a particle, operating at 1.8 GHz, is shown in Fig.~\ref{fig:omega} and the corresponding scattering directivity pattern in Fig.~\ref{fig:omega2}(a). The dimensions are the same as for the chiral particle, except that the separation between the dipole arms (i.e., the width of the loop gap) is 2.8~mm. The distance between the two loops is 1~mm. In this case, the forward scattering is clearly minimized and the pattern has a similar shape to the ideal pattern of Fig.~\ref{fig:1}, although the maximum directivity is slightly below 3. The scattering directivity in the forward direction is as small as 2.3$\times 10^{-7}$. The corresponding axial ratio pattern (absolute values) is shown in Fig.~\ref{fig:omega2}(b). The axial ratio is 0 (i.e., the scattered wave has linear polarization) for almost all the scattering directions. It should be noted that in order to have a truly uniaxial particle, another identical element should be added, orthogonal to the existing one, similarly to the uniaxial chiral particle considered above. The second element has been left out in this case for clarity and because it is practically invisible to the linearly polarized incident field as it is defined here.

\section{Discussion and Conclusions}

In this paper, scattering of a linearly polarized plane wave by
small uniaxial bi-anisotropic particles has been studied. Firstly,
the conditions for zero backward, forward, and total scattering
conditions have been derived for a general uniaxial bi-anisotropic
particle with all the polarizability dyadics having both symmetric
and antisymmetric components. Zero backscattering as well as zero
forward scattering for oblique incidence was shown to be possible
only for certain special cases under the assumption that the axial polarizabilities are zero. For the axial incidence, requiring
zero backscattering or zero forward scattering lead in both cases to
two independent equations: one between the antisymmetric parts of
the self-coupling polarizabilities and symmetric parts of the
cross-coupling polarizabilities and one between the symmetric parts
of the self-coupling polarizabilities and antisymmetric parts of the
cross-coupling polarizabilities. Having zero total scattering from a
particle, i.e., making the particle invisible, for one axial
incidence direction was shown to be possible by balancing the
particle response in particular ways. Secondly, four special cases
of the general uniaxial bi-anisotropic particle (omega, moving
particle, chiral, and Tellegen particles) have been analyzed and
discussed in detail. The results show that zero backscattering
property can be achieved for the axial excitation of a chiral or
moving particle as long as the condition
$\alpha_\lt{ee}=\frac{1}{\eta_0^2}\alpha_\lt{mm}$ is satisfied. A
more complex condition, not satisfied by a ``conventional'' omega
particle, of $-\alpha_\lt{ee} \pm\frac{2}{\eta_0}\alpha
+\frac{1}{\eta_0^2}\alpha_\lt{mm}=0$ where $\pm$ corresponds to the
opposite directions of the axially incident plane wave is required
for an omega particle whereas zero backscattering is not possible
for a Tellegen particle. The scattering pattern is the same in all
the cases though the scattered wave has different polarizations.
Zero forward scattering, on the other hand, can be achieved only for
incident waves propagating along the particle axis with an omega or
a Tellegen particle by fulfilling the condition
$\alpha_\lt{ee}=-\frac{1}{\eta_0^2}\alpha_\lt{mm}$ or with a moving
particle by fulfilling the condition $-\alpha_\lt{ee} \pm
\frac{2}{\eta_0}\alpha - \frac{1}{\eta_0^2}\alpha_\lt{mm}=0$ where
$\pm$ again correspond to the opposite directions of the axially
propagating incident plane wave. Zero forward scattering condition
is not possible to satisfy for a chiral particle. Moreover, the zero forward
scattering condition can only be satisfied exactly with an active particle. The
scattering pattern is the same for all the cases, but the
polarization of the scattered wave differs. Zero total scattering
for one incident direction along the particle axis can be achieved
for a moving particle by satisfying $\alpha_\lt{ee} = \pm 1/\eta_0
\alpha= 1/\eta_0^2 \alpha_\lt{mm}$ or for an omega particle by
satisfying $\alpha_\lt{ee} = \pm 1/\eta_0 \alpha= -1/\eta_0^2
\alpha_\lt{mm}$ where the top sign corresponds to the positive
propagation direction. When the illuminating plane wave travels in
the opposite direction, the particle does not scatter in the
backward direction in the case of an omega particle or the forward direction in the case of a moving particle, but does scatter in all the other directions.

The conditions for zero backscattering,  zero forward scattering and
zero total scattering when the scatterer is an omega particle, a
moving particle, a chiral particle and a Tellegen particle are
summarized in Table~\ref{T:2}. The axial ratio of the scattered wave
when the corresponding scattering condition is met (the axial
incidence with linear polarization) is also shown for each particle.
Interestingly, we can see a kind of symmetry in the required
conditions, as the conditions are the same for a chiral particle and
a moving particle for zero backscattering as well as for an omega
particle and a Tellegen particle for zero forward scattering.  It
should also be noted that all the results are in agreement with the
zero backscattering conditions derived earlier in \cite{Wagner1963,
Weston1963, Lindell2009} and with the optical theorem.

Although in some of the studied cases shown in Table~II the required particle can be easily realized as was demonstrated here numerically with zero-backscattering chiral particle and zero forward scattering omega particle, the practical
design and in some cases even the feasibility of the required
particle are as of yet unknown for many particle types, especially for the nonreciprocal ones. Conceptual discussions of possible
structures exhibiting all the types of bi-anisotropic coupling can
be found, e.g., in \cite{Serdyukov2001}.

\end{document}